# Coherent plasmon and phonon-plasmon resonances in carbon nanotubes

Abram L. Falk[1,*], Kuan-Chang Chiu[1,2], Damon B. Farmer[1], Qing Cao[1], Jerry Tersoff[1], Yi-Hsien Lee[2], Phaedon Avouris[1], Shu-Jen Han[1]

[1]IBM T. J. Watson Research Center, Yorktown Heights, NY 10598, USA.
[2]Dept. of Materials Science and Engineering, National Tsing-Hua University, Hsinchu, Taiwan.

## Abstract

Carbon nanotubes provide a rare access point into the plasmon physics of one-dimensional electronic systems. By assembling purified nanotubes into uniformly sized arrays, we show that they support coherent plasmon resonances, that these plasmons enhance and hybridize with phonons, and that the phonon-plasmon resonances have quality factors as high as 10. Because coherent nanotube plasmonics can strengthen light-matter interactions, it provides a compelling platform for surface-enhanced infrared spectroscopy and tunable, high-performance optical devices at the nanometer scale.

## Main text

Plasmons in carbon nanotubes [1–5] comprise longitudinal charge oscillations coupled to infrared or terahertz optical fields. They can either propagate [3–5] or be confined to Fabry–Pérot resonators by reflections at the nanotube ends [6–10] (Fig. 1(a)). Propagation losses are low [3], and the resonant frequencies and absorption coefficients can be controlled via the length [8,10] and doping level [7,9] of the nanotubes. The intense concentration of electromagnetic fields deriving from the nanotubes' one dimensionality allows the plasmons both to confine light to the nanometer scale and to enhance light-matter interactions by Purcell factors that are predicted to be as high as $10^6$ [5].

At infrared frequencies, nanotube plasmonics could lead to highly sensitive absorption spectroscopy through surface enhanced infrared absorption (SEIRA) [11–13]. At terahertz frequencies, it could enable tunable lasers and receivers for use in terabit-per-second wireless communications [14–16]. Ultra-broadband nanotube plasmonic circuitry could be naturally integrated with high-performance nanotube transistors.

However, nanotube plasmonics has been frustrated by the material quality of nanotube films. In inhomogeneous nanotube films, with a broad distribution of lengths, diameters and/or doping levels, plasmons resonating at different frequencies quickly lose phase coherence with each other, leading to fast dissipation. The quality ($Q$) factor, which is the quotient of the resonant angular frequency and the dissipation rate, is therefore low. To date, this inhomogeneity has limited observations of nanotube plasmon resonators to the incoherent $Q \ll 1$ regime [6–10]. Because dissipation constrains nearly all applications of plasmonics, the demonstration of high $Q$ resonators would provide crucial evidence that nanotubes are a technologically viable plasmonic material.

---
* Contact: alfalk@us.ibm.com

In this work, we show that coherent nanotube plasmon and phonon-plasmon resonances can have ensemble $Q$ factors as high as 10. The key to our demonstration is our exceptionally uniform nanotube films, which we develop using Langmuir-Schaeffer techniques [17]. We conservatively estimate that our nanotube resonators confine an electromagnetic field whose free-space wavelength ($\lambda_0$) is 8 μm to a mode volume ($V$) of 0.002 μm$^3$. With this combination of $Q$ and optical concentration ($\lambda_0^3/V = 300{,}000$), the Purcell factor by which these plasmonic resonators could enhance light matter interactions [18], $P = (3Q/4\pi^2)(\lambda_0^3/V)$, exceeds 100,000. As an illustration of a key technology arising from these extraordinary Purcell factors, we demonstrate nanotube-based SEIRA spectroscopy. All of the assembly and fabrication processes developed in this work can be readily extended to the wafer scale.

The supporting substrates for our carbon nanotube films are Si with 10 nm of SiO$_2$ and either 0, 10, or 40 nm of diamond-like carbon (DLC) on top of the SiO$_2$. The DLC is a non-polar spacer that controls the coupling between plasmons in the nanotubes and polar phonons in the SiO$_2$. To assemble the nanotube films, we added 99.9% semiconducting nanotubes in 1,2-dichloroethane to the wafer surface of a Langmuir trough. Mechanical moving bars compressed the suspension to a target pressure of 30 mN m$^{-1}$ with a bar moving rate of 20 cm$^2$ min$^{-1}$ under multiple (~10) isothermal cycles. We transferred the 6-nm thick nanotube arrays onto the various target substrates and used oxygen-based reactive-ion etching to cut the nanotubes into short segments, whose length ($L$) ranged from 30 to 500 nm (Fig. 1(b)). We then doped them to be p-type via exposure to NO$_2$ gas [19] and studied the plasmon resonances with micro Fourier-transform infrared spectroscopy (μ-FTIR) at room temperature. Each measurement spatially averages 1-10 million nanotubes in a ~(25 μm)$^2$ area.

We observe three prominent resonances from the nanotube segments (Fig. 2(a)). The frequency of the highest frequency resonance ($\nu_1$) conspicuously increases as the segment length ($L$) decreases, a behavior that identifies $\nu_1$ as the Fabry–Pérot resonance of the nanotube plasmon. At $L$= 30 nm, $\nu_1$ reaches 3000 cm$^{-1}$, corresponding to a free space wavelength of $\lambda_0 = 3$ μm. This result shows that nanotube plasmon resonances, previously observed at THz through far-infrared frequencies [6–10], can also span the mid-infrared frequency range.

When we lithographically cut the nanotube films parallel to the nanotube alignment direction, not perpendicular to it, $\nu_1$ is broader and both $\nu_1$ and $\nu_3$ are much weaker (Fig. 2(b)). This measurement confirms that uniform segmentation (i.e. length control) of the nanotubes is important to the coherence of their ensemble plasmon resonance. Moreover, the attenuation of light polarized parallel to the nanotube alignment direction ($A_\parallel$) is much stronger than that of perpendicularly polarized light ($A_\perp$), with $A_\parallel / A_\perp = 6 \pm 1$ (Fig. 2(b), inset). The observed ratio indicates good, though imperfect, nanotube alignment.

A distinctive feature of nanotube plasmonic resonators is that their doping level can dynamically control the resonant frequencies and amplitudes. Electrically gated nanotube plasmons could therefore lead to modulators for THz-frequency communications and tunable SEIRA spectrometers. Figure 2(c) shows that the doping level (induced by $NO_2$ adsorption) strongly affects the plasmon resonances, with a highly p-type doped film exhibiting a 6 times stronger $\nu_1$ attenuation peak than same film when undoped. Doping also causes $\nu_1$ to blueshift from 1370 cm$^{-1}$ to 1920 cm$^{-1}$, a 40% change in frequency.

The frequency of the lower energy resonances ($\nu_2$ and $\nu_3$) evolve much more weakly with $L$ than $\nu_1$ does. However, their intensity and shape is a strong function of $L$, and in particular, are strongly enhanced as the frequency of $\nu_1$ approaches their frequencies. These behaviors indicate that they derive from plasmon-coupled phonons. By comparing the nanotubes' spectra to spectra of graphene nanoribbons on $SiO_2$, we identify $\nu_3$ with the 1168 cm$^{-1}$ the longitudinal optical (LO) phonon in the $SiO_2$ substrate [16,20]. The $\nu_2$ resonance at 1590 cm$^{-1}$ corresponds to the infrared-active $E_1$ and $E_{1u}$ phonon modes [21,22] of carbon nanotubes. These phonon modes, though closely related to the strong G-band Raman modes of nanotubes [23], are ordinarily very weak spectroscopic features.

However, plasmon coupling significantly strengthens them and modifies their shape (Fig. 3). For $L <$ 60 nm, when $\nu_1$ is strongly off resonance with 1590 cm$^{-1}$, $\nu_2$ is absent. As $L$ increases, the $\nu_2$ feature appears as an attenuation peak. As $L$ continues to increase, its lineshape becomes asymmetric and then evolves into a window of phonon-induced transparency. The intensity of $\nu_2$ also evolves with the nanotube doping level and is much stronger when $\nu_1$ is on resonance (Fig. 2(c)). This resonant enhancement demonstrates the principle of SEIRA spectroscopy and could be used to sensitively detect and probe external chemicals like gases or biomolecules.

The intensity of $\nu_3$, the $SiO_2$ phonon resonance, also strongly increases as $L$ increases and $\nu_1$ approaches $\nu_3$. Moreover, as can be seen in Fig. 2(a), while the lineshape of $\nu_1$ is symmetrical for the shortest nanotubes, it evolves into an increasingly asymmetrical lineshape as $\nu_1$ approaches $\nu_3$. Understanding the lineshapes as a result of Breit-Wigner-Fano interference [24,25] between plasmon and plasmon-coupled phonon resonances, we fit all of the of $\nu_1$ and $\nu_2$ lineshapes to:

$$A(\nu) \propto \frac{(F_j \gamma_j + \nu - \nu_j)^2}{(\nu - \nu_j)^2 + \gamma_j^2} + b_j \quad (1)$$

with $\gamma_j$ the linewidth, which is the damping rate divided by $2\pi$, and $F_j$ the Fano parameter, which characterizes the resonance's degree of asymmetry, $b_j$ an offset parameter that represents screening of the resonance by intrinsic losses [26], and $j = 1$ or 2. These fits exhibit excellent agreement with the measured spectra (Figs. 2(a), 3(a) and 3(b)). We do not fit $\nu_3$ lineshapes in this work, because this peak, though primarily deriving from the 1168 cm$^{-1}$ phonon in $SiO_2$, also shows interference from the 806 cm$^{-1}$ longitudinal optical phonon in $SiO_2$. We cannot derive all of the rates describing the coupled phonon-plasmon coupling from our measured spectra alone, and the crossover from electromagnetically induced transparency (EIT) to the strong coupling / vacuum Rabi splitting regime is complex [27,28]. Nevertheless, because $\nu_2$ is a much sharper feature than that of $\nu_1$, we interpret it as a phonon-induced transparency feature. On the other hand, the plasmon-SiO$_2$ phonon interaction is at the threshold of the strong coupling regime, as we will show below.

When plotting $\nu_1$ and $\nu_3$ vs. wavevector ($q$), where $q = \pi/L$, we observe a clear anticrossing (Fig. 4(a)). In the short $L$ (high $q$) limit, $\nu_1$ is asymptotically linear with $q$, consistent with the theory of plasmon resonances of one-dimensional electronic systems [1–3,29,30]. The slope of this linear asymptote in Fig. 4a (0.0032) implies an asymptotic effective index of refraction of $1/(2\pi \times 0.0032) = 50$ and a plasmon velocity ($V_p$) of $c / 50$. In terms of the Fermi velocity ($V_F = c / 300$), $V_p = 6V_F$.

The $\gamma_1$ linewidths and corresponding $Q$ factors (Fig. 4(b)-(c)) are the key metrics of coherence. When $L > 200$ nm, $\gamma_1^{-1} = \sim 200$ fs. As $L$ decreases, $\gamma_1^{-1}$ decreases, reaching 50 fs for $L = 100$ nm and 20 fs for $L = 40$ nm. Calculating $Q$ as $\nu_1/\gamma_1$, we find that $Q$ has a maximum of 10 and also decreases with decreasing $L$ until $L < 100$ nm, at which point it remains an approximately constant $Q = 2$. Among plasmon resonances with minimal coupling to SiO$_2$ phonons, $Q = 3$ was the highest quality factor that we observed. The $Q$ vs. $L$ trend derives from multiple factors, including lithographic effects, nanotube alignment effects, an increased number of phonon decay pathways for higher energy plasmons [16], and phonon-plasmon hybridization near $\nu_3$. Coupled antenna theory also indicates that mutual coupling in dense nanotube arrays blue-shifts and broadens ensemble plasmon resonances [31].

The frequency of the $\nu_1$, $\nu_3$ splitting at the 1168 cm$^{-1}$ anticrossing corresponds to twice the plasmon-phonon coupling strength ($g_{13}$, see Fig. 4(d)). As expected, $g_{13}$ is highest (6.5 THz) when the nanotubes rest directly on the SiO$_2$. This coupling strength is comparable to the ~5 THz linewidth of $\nu_1$ and $\nu_3$ near their anticrossing, indicating that the $\nu_1$-$\nu_3$ interaction is at the threshold of the strong coupling regime. Fitting $g_{13}$ as a function of the DLC-spacer thickness ($x$) to $e^{-x/\delta}$ provides an estimate of the transverse decay length of the plasmon mode (Fig. 4(e)). This length, $\delta = 55$ nm, is 150 times smaller than $\lambda_0 = 8.3$ μm (1200 cm$^{-1}$) at $L = 180$ nm. We estimate the maximum Purcell factor at the mode center to be $P = 180,000$ [32]. Isolated nanotubes are predicted to support even much higher $P$ factors than this extremely high value [5,33].

The plasmon resonances of graphene nanoribbons [16,34–37] share several features with those that we observe in nanotubes, including plasmon-phonon hybridization [16] and phonon-induced transparency [35]. However, nanotube plasmonics has several unique prospects. The one dimensionality of nanotubes allows them to confine light to smaller volumes and to suffer from less scattering at edges [4,5]. Their linear dispersion relationship, compared to the square-root relationship for graphene [16,34], makes their resonances more broadly tunable. Highly absorptive thick films of nanotubes would have no clearly manufacturable analogue in graphene. Finally, compared to those of metallic graphene, the plasmon resonances of semiconducting nanotubes have both stronger doping effects and a potential to be more efficient in photothermoelectric detectors [38].

In conclusion, we showed that carbon nanotubes can be a high quality plasmonic nanomaterial that strengthen light-matter interactions. As demonstrated by the resonant enhancement of phonons that we observe, nanotube plasmonics should be an ideal tool for sensitive SEIRA spectroscopy. In this application, nanotubes could play a dual role as plasmonic resonators and infrared receivers, and functionalization could enhance their interaction with specific analytes. Their nonlinearity and anisotropic optical properties also make them promising building blocks for metamaterials [39]. In the long run, because nanotubes can function both as electrical transistors and nanophotonic modulators, they could be a foundation for integrating electrical and optical logic at the nanometer scale.

**Acknowledgments**

The authors thank Hendrik Hamann, Easwar Magesan, David Toyli, James Hannon, and Jianshi Tang for discussions, and Jim Bucchignano for help preparing samples. This work was funded by IBM.

**Figures**

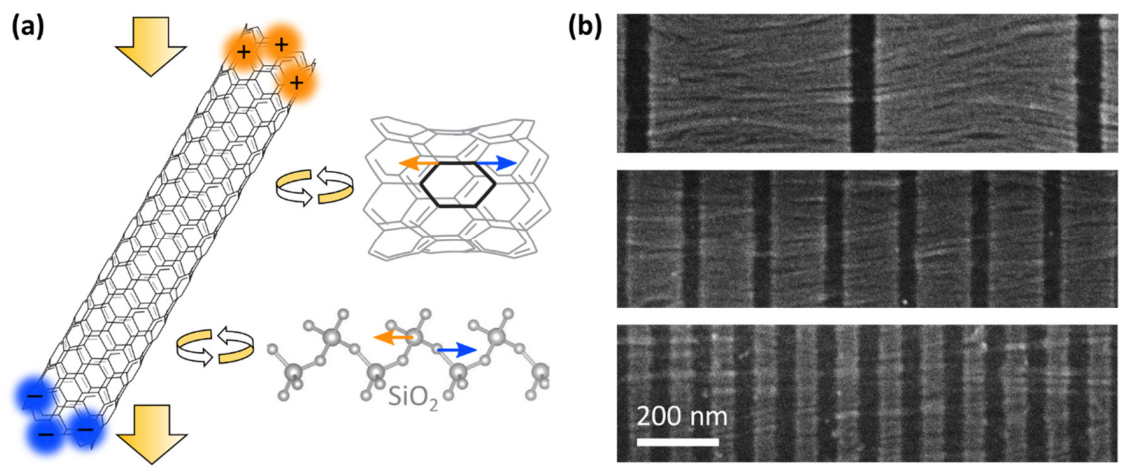

FIG. 1. (a) Finite-length nanotubes act as plasmonic Fabry–Pérot resonators. The plasmon resonances are excited with broad-spectrum infrared light in a µ-FTIR spectrometer. Before decaying, they can couple to phonons in the nanotube or phonons in the $SiO_2$ substrate. (b) Scanning electron micrographs of cut and aligned nanotube segments. The nanotube film thickness is 6 nm.

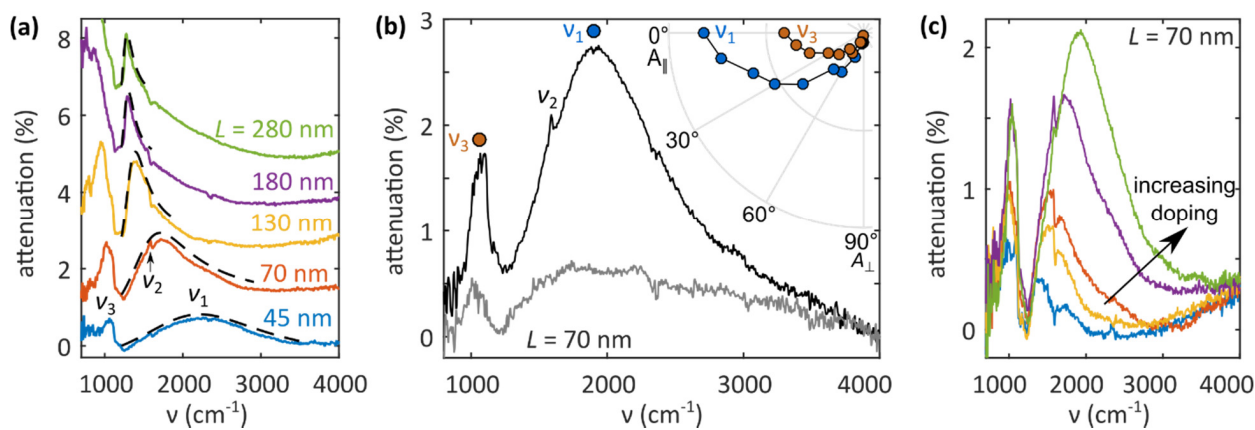

FIG. 2. (a) The attenuation spectra of nanotube resonators when directly resting on the $SiO_2$ substrate, showing both plasmon ($\nu_1$) and plasmon-coupled phonon (($\nu_2$ and $\nu_3$) resonances. The spectra are sequentially vertically offset, for clarity, and the black dashed lines are fits to Fano functions. (b) Cut-direction dependence of the plasmonic absorption. The black (gray) curve is the attenuation spectrum of nanotubes cut perpendicular (parallel) to their alignment direction. Inset: polarization dependence of the intensity of peaks $\nu_1$ and $\nu_3$, with the cut direction fixed to be the perpendicular direction (as shown in Fig. 1(b)). The intensities of peaks $\nu_1$ and $\nu_3$ are each $6 \pm 1$ times greater when the polarizer is oriented along the nanotube length rather than perpendicular to it. (c) Doping dependence of the nanotube segments. The green curve is measured immediately after $NO_2$ exposure. For each curve below it, the samples have rested in atmospheric conditions for one extra day, whereby the doping level decreases.

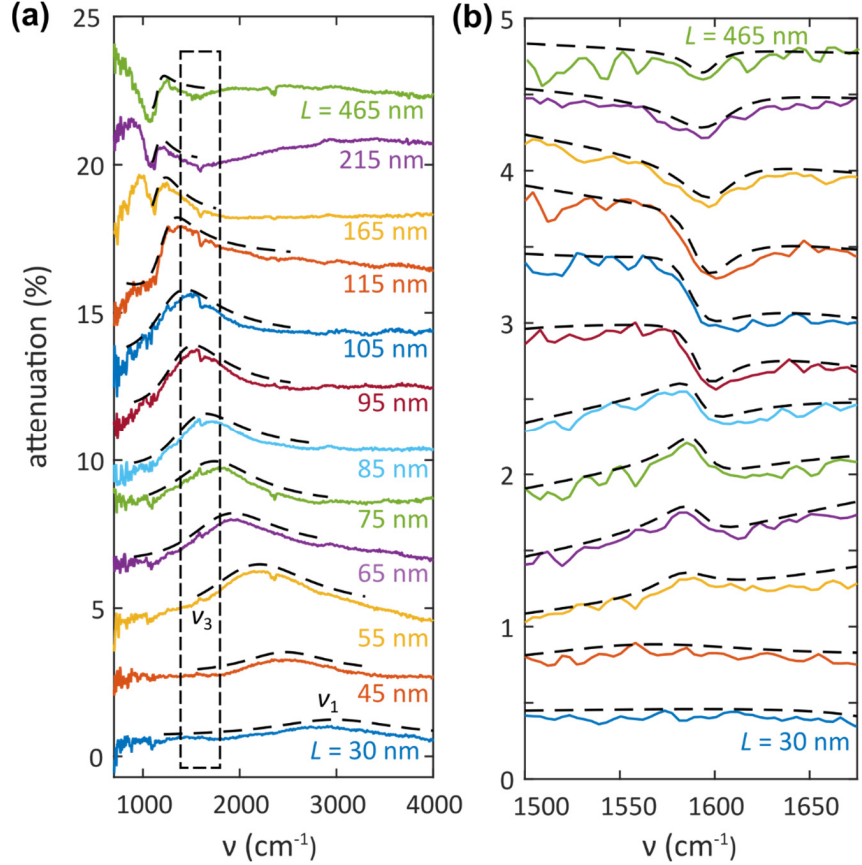

FIG. 3. (a) The attenuation spectra of nanotube segments on the 40-nm-DLC substrate, showing $\nu_1$ decrease with increasing $L$ and pass through the 1590 cm$^{-1}$ nanotube phonon resonance. The spectra are sequentially vertically offset, for clarity, and the black dashed lines are fits to Fano functions, which are also offset from the data. (b) A magnification of the dashed rectangle from Fig. 3(a), along with fits to Fano functions (black dashed lines), showing the 1590 cm$^{-1}$ $E_{1u}$ nanotube phonons when $\nu_1$ is on or nearly on resonance. As $L$ increases and $\nu_1$ moves through the 1590 cm$^{-1}$ resonance, the character of the 1590 cm$^{-1}$ phonon resonance changes from being absent, to an absorbing feature (due to constructive interference), to a partial transparent feature (due to destructive interference). The fitted resonance width ($\gamma_2$) ranges from 20 – 30 cm.

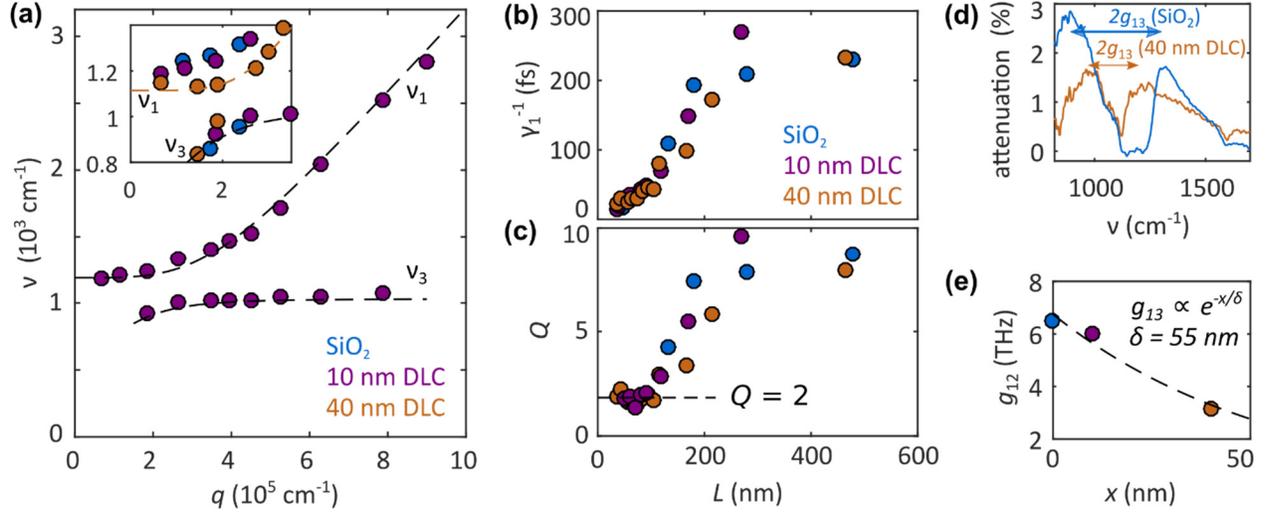

FIG. 4. (a) The resonant frequencies, $\nu_1$ and $\nu_3$, as a function of wave vector ($q = \pi/L$). The dashed black lines are guides to the eye. The $\nu_1$ frequencies derive from fits, and the $\nu_3$ frequencies derive from selecting the frequency corresponding to the local maximum of the attenuation spectra. Inset: $\nu_1$ and $\nu_3$ vs. $q$ near the anticrossing at 1168 cm$^{-1}$. The splitting between $\nu_1$ and $\nu_3$ is smaller when the DLC spacer is larger. (b) $\gamma_1^{-1}$, derived from fits to Equation (1) as a function of $L$. (c) $Q$, computed as $\nu_1/\gamma_1$, as a function of $L$. (d) The attenuation spectrum at the anticrossing ($L = 180$ nm, $q = 1.8 \times 10^5$ cm$^{-1}$), showing a higher $g_{13}$ splitting for the nanotubes resting on SiO$_2$ compared to those resting on the 40-nm DLC substrate. (e) $g_{13}$ as a function of the DLC thickness ($x$).

# Supplemental Material for "Coherent plasmon and phonon-plasmon resonances in carbon nanotubes"

Abram L. Falk[1], Kuan-Chang Chiu[1,2], Damon B. Farmer[1], Qing Cao[1], Jerry Tersoff[1], Yi-Hsien Lee[2], Phaedon Avouris[1], Shu-Jen Han[1]

[1]IBM T. J. Watson Research Center, Yorktown Heights, NY 10598, USA.
[2]Dept. of Materials Science and Engineering, National Tsing-Hua University, Hsinchu, Taiwan.

**Contents:**

**S1. Atomic force microscopy**
**S2. Fano parameter for $\nu_2$**
**S3. Fabrication of homogeneous nanotube films**
**S4. Spectroscopy**
**S5. Purcell Factor calculation**
**S6. Benzyl-viologen doping**

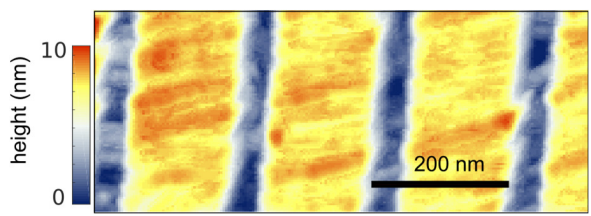

FIG S1. An atomic force micrograph of a fabricated nanotube array. The thickness of the nanotube film is 6 nm.

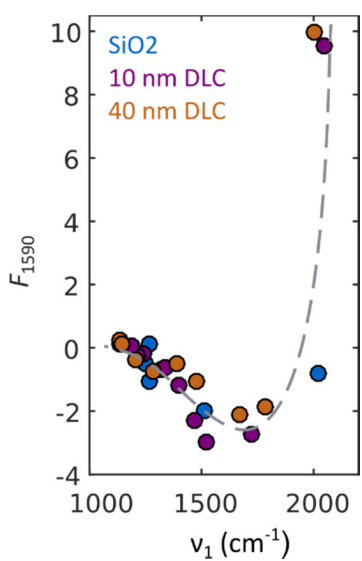

FIG S2. The $F_2$ Fano parameter of the 1590 cm$^{-1}$ $\nu_2$ feature, as a function of $\nu_1$, when fitting the resonances in Fig. 3(b) to Equation 1 in the main text. As expected from the Fano model, the asymmetry parameter ($F_2$) is $\gg 1$ when the plasmon-phonon interference is symmetric and constructive, near 0 when the interference is asymmetric and destructive, and at intermediate negative values for the other cases. The dashed line is a guide to the eye.

## S3. Fabrication of homogeneous nanotube films

Substrates were fabricated at the Materials Research Laboratory at IBM's T.J. Watson Research center. Alignment marks were etched into 8-inch silicon wafers and 10 nm of $SiO_2$ was thermally grown on top. Plasma enhanced chemical vapor deposition was used to deposit the DLC spacer layer. The wafers were then diced into 2.5 inch chips.

A Langmuir-Schaefer technique was used to assemble the nanotube films [17]. 0.3 mg of single-walled carbon nanotube powder with semiconducting nanotube purity of 99.9% (IsoNanotube-S, Nanointegris) was dispersed in 15 mL of 1,2-dichloroethane (DCE, >99.0%, Sigma-Aldrich), together with 1 mg of poly(p-phenylenevinylene-co-2,5-dioctyloxy-m-phenylenevinylene) (PmPV, Sigma-Aldrich). This suspension was then sonicated with a high-power horn sonicator (60 min, 600 W, 95% amplitude, 20 kHz). Centrifugation (210,000 g for 2 h) then removed the sediment. The nanotube suspension was filtered through a polytetrafluoroethylene filter membrane (Fluoropore-Millipore Membrane, pore size 0.2 µm, 25 mm) and then washed with clean DCE to remove excess PmPV. The nanotubes were then redispersed on the filter paper in 15 mL of clean DCE through a brief bath sonication.

To assemble these dispersed semiconducting nanotubes into aligned arrays, 50 µl of the nanotube DCE suspension was added to the water surface of a Langmuir trough (Model 611, Nima Technologies). The suspension was then compressed to a target pressure of 30 mN m$^{-1}$ with a bar moving rate of 20 cm$^2$ min$^{-1}$ under multiple (~10) isothermal cycles. The nanotube arrays were then transferred onto the various target substrates. Atomic-force microscopy showed that the films are 6-nm thick.

To cut the nanotubes into segments of uniform length, an etch mask on the chips was fabricated with conventional electron beam lithography and a poly(methyl methacrylate) (PMMA) mask. The nanotubes were etched with one minute of oxygen reactive ion etching and the PMMA etch mask was then removed with acetone. The chips were then annealed in a vacuum oven at 450 degrees for thirty minutes, to burn off any remaining organics. To dope the nanotubes, flowing $NO_2$ gas pulses were dosed at 25°C under a vacuum of 300 mTorr for 5 min.

## S4. Spectroscopy

The attenuation spectra were measured in a Bruker Nicolet 8700 µ-FTIR system. The diameter of the focused beam size is 25 µm, and each measurement was integrated for a 30-second period. Immediately before capturing each transmittance spectrum ($T$), a background spectrum ($T_0$) on an area of the chip without nanotubes was measured. The attenuation spectrum is computed as attenuation = 1-$T$/$T_0$.

For most of the measurements in this manuscript, a polarizer in our apparatus was not used. When the light was polarized in the alignment direction of the nanotubes (inserting a polarizer between the sample and the detector), a significant change in the spectra was not observed, aside from a lower signal to noise due to the polarizer's attenuation. This similarity in shapes is expected from a sample patterned with aligned nanotubes, which creates a highly anisotropic dielectric constant in the plane of the nanotube film. To obtain the polarization-dependent results in Fig. 2(b), the direction of the polarizer was rotated. To confirm that our polarization results did not stem from an overall anisotropy of the incident light or detector, the direction of the polarizer was also left constant and the sample was rotated. In this case, similar results to those when the polarizer was rotated were observed.

## S5. Purcell factor calculation

In the weak coupling regime, an emitter in a cavity will have its emission rate modified due to the difference in the cavity's optical density of states relative that of free space. For an ideally situated dipole, the Purcell factor ($P$) describing the degree of this modification is:

$$P = \frac{3Q}{4\pi^2}\left(\frac{\lambda_0^3}{V}\right), \tag{S1}$$

where $Q$ is the cavity's $Q$ factor, $\lambda_0$ is the free space wavelength, and $V$ is the mode volume [18]. We estimate the transverse extent of the nanotube plasmon mode ($\delta$) as the characteristic $1/e$ decay of $g_{12}$, the coupling strength between the plasmon and the 1000 cm$^{-1}$ SiO$_2$ phonon, as the spacing between the SiO2 and the nanotubes is increased. At the $\nu_1$, $\nu_2$ anticrossing, $\delta = 55$ nm, $L = 180$ nm, and $Q = 7$. We then approximate the plasmon mode as a cylinder and calculate $V$ as $V = \pi\delta^2 L$. We obtain $V = .0017$ µm$^{-3}$, $V/\lambda_0^3 = 3 \times 10^{-6}$. Using Equation (2), we then calculate an ideal Purcell factor of $P = 180,000$ at the center of the mode. We emphasize that these estimates are not intended to be precise, as they exclude coupling between nanotubes and mode distortion due to dielectric constant of the substrate.

Purcell factors from ideal, isolated nanotubes are predicted to be much higher than even this very high value [3,5]. The $\delta$ that we measure should be understood as a property of nanotubes in our actual film. For a variety of reasons, $\delta$ can be larger for nanotubes in a dense film than it is for single isolated nanotubes [4]. Moreover, the $Q$ factor of an isolated nanotube will likely be higher than that of nanotubes in an ensemble. Finally, near the phonon-plasmon anticrossing where we measure $\delta$, the nanotube plasmon has hybridized with the phonon, which modifies its spatial extent. Ref. 18 shows that for, a plasmon mode on the surface of a metal cylinder of radius $a$, $\delta$ will be comparable to $a$ in the quasistatic limit ($a \ll \lambda_0$). In Ref. [3], electromagnetic simulations calculate that $\delta$ for nanotubes is ~ 1 nm (at $\lambda_0 = 10.6$ µm). In Ref. [5], $P$ exceeding $10^6$ for dipoles coupled to traveling plasmons in nanotubes is predicted. In a nanotube resonator, $P$ would be even higher.

## S6. Purcell factor calculation

As explained in the main text, we doped the nanotubes with adsorbed $NO_2$ molecules in order to add positive charge carriers to the nanotubes, which blueshifts and strengthens the plasmon resonance. We also attempted to dope the nanotube n-type by spinning a PMMA / benzyl viologen (BV) composite on top of the nanotubes. This composite is known to make nanotube transistors have an n-type gate transfer curve. We observed it to redshift the plasmon resonance (Fig. S3). The likely reason for this redshift is that because our nanotubes are not wired up into a transistor configuration, the n-type carriers are not fully activated, and serve to simply compensate for adsorbed water on the nanotubes, which is known to induce positive charge carriers in nanotubes. If the n-type charge carriers were fully activated, n type doping would blueshift the plasmon resonance, much as p-type doping does.

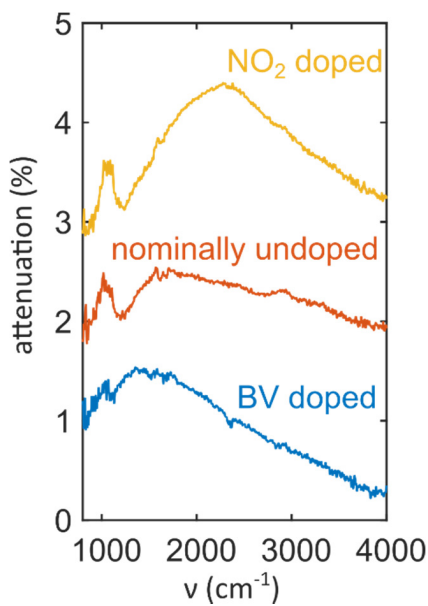

FIG S3. Attenuation spectra of 110-nm-long nanotubes under three different doping configurations: 1. Doping with $NO_2$ adsorption (positive charge carriers). 2. Unintentional doping (positive charge carriers due to adsorbed water molecules). 3. Benzyl viologen (BV) partially compensates for the unintentional p-type doping, thereby redshifting the plasmon resonance.